\documentclass[sigconf]{acmart}

\usepackage{adjustbox}
\usepackage{amssymb}
\usepackage{bm}
\usepackage{booktabs}
\usepackage{caption}
\usepackage{color}
\usepackage{makecell}
\usepackage{microtype}
\usepackage{multirow}
\usepackage{paralist}
\usepackage{subcaption}
\usepackage{tabularx}
\usepackage{url}
\usepackage{xcolor}

\usepackage{balance}

\usepackage{amsmath}
\usepackage{amsfonts}
\usepackage{dsfont}
\usepackage{xfrac}
\usepackage{subcaption}
\usepackage{caption} 
\PassOptionsToPackage{hyphens}{url}\usepackage{hyperref}
\definecolor{mylinkcolor}{RGB}{0,0,0}
\hypersetup{colorlinks,allcolors=mylinkcolor,citecolor=mylinkcolor}
\usepackage[capitalize]{cleveref}
\newcommand{\xhdr}[1]{\vspace{1.0mm}\noindent{{\bf #1.}}\hspace{0.5mm}}

\begin{document}
\title{Choosing to grow a graph: \\ Modeling network formation as discrete choice}

\author{Jan Overgoor}
  \affiliation{\institution{Stanford University}}
  \email{overgoor@stanford.edu}
\author{Austin R. Benson}
  \affiliation{\institution{Cornell University}}
  \email{arb@cs.cornell.edu}
\author{Johan Ugander}
  \affiliation{\institution{Stanford University}}
  \email{jugander@stanford.edu}

\begin{abstract}
We provide a framework for modeling social network formation
through conditional multinomial logit models from discrete choice and random utility theory, in which
each new edge is viewed as a ``choice'' made by a node to connect to another node,
based on (generic) features of the other nodes available to make a connection.
This perspective on network formation unifies
existing models such as preferential attachment, triadic closure, and node fitness, which are all special cases,
and thereby provides a flexible means for
conceptualizing, estimating, and comparing models.
The lens of discrete choice theory also provides several new tools for analyzing social network formation;
for example, the significance of node features can be evaluated in a statistically rigorous manner, and mixtures of existing models can be estimated by adapting known expectation-maximization algorithms.
We demonstrate the flexibility of our framework 
through examples that analyze a number of synthetic and real-world datasets.
For example, we 
provide rigorous methods for estimating preferential attachment models and show how to separate the effects of preferential attachment and triadic closure. Non-parametric estimates of the importance of degree show a highly linear trend, and we expose the importance of looking carefully at nodes with degree zero. Examining the formation of a large citation graph, we find evidence for an increased role of degree when accounting for age.
\end{abstract}

\maketitle

\section{Introduction}

Understanding how networks form and evolve
is an essential component of understanding their structure,
which in turn underlies the basis for understanding the 
broad range of processes that occur {\it on} networks.
Models of social network formation can largely be decomposed into
node formation and edge formation.
In this work, we argue that edge formation can be effectively modeled as a \emph{choice} made by an actor (or actors) in the network to instantiate a connection to another node.
The diverse research on network formation has led to many models and mechanisms of edge formation, including 
  preferential attachment \cite{albert99},
  uniform attachment \cite{callaway01},
  triadic closure \cite{jin01},
  random walks \cite{vazquez03,saramaki04},
  homophily \cite{papadopoulos12},
  copying edges from existing nodes \cite{kleinberg99,kumar99,kumar00},
  latent space structures \cite{hoff02,leskovec07,papadopoulos12},
  inherent node fitness \cite{caldarelli02},
  and combinations of all of these \cite{liu02,jackson07,leskovec08}.
Here, we frame edge formation as a discrete choice process
and derive a family of discrete choice models
\cite{mcfadden73,train09}
that subsume a wide range of existing models in a 
unified framework and also naturally opens up
a host of powerful extensions.

Discrete choice models are commonly employed in economics,
social psychology, and statistics as a way to model how individuals make choices from
a slate of discrete alternatives \cite{agresti03}.
Typically, the alternatives have associated features, and statistical
models of discrete choice make it possible to estimate the relative
importance of such features.
Such models have been used to answer questions such as
  how consumers choose goods \cite{simonson92},
  how people choose where they live \cite{mcfadden78},
  how students choose what college to attend \cite{fuller82},
  and how commuters choose between different modes of transportation \cite{train78}.
Discrete choice analysis is also used to understand how choices vary depending on the context in which they are framed:
in online commerce, this could be how web layouts lead to different purchasing priorities \cite{ieong12};
for choosing colleges, this could be incorporating the effect of the national economy.
In this paper, we demonstrate how discrete choice models can similarly help
us understand the factors driving social network evolution.
 
The starting point for the present work is the observation that edge formation events in social networks are naturally viewed as discrete choices.
For simplicity, consider a directed graph where edges are formed one by one,
where we can think of the formation of a directed edge $(i, j)$
as $i$ ``choosing'' to connect with $j$, 
where the set of alternatives available to $i$ is the set of all other nodes.
(While undirected graph models are common in social network analysis,
the underlying formation procedure is almost always asymmetric. For example,
the Facebook friendship graph is typically modeled as an undirected graph~\cite{ugander11},
but the friendships are proposed by one of the two nodes in an edge.)
The key modeling question is easy to state: why did $i$ choose $j$?
This question has long been the informal subject of network formation modeling and at the same time the exact question that discrete choice models and analysis have been designed to answer.
However, up to this point, network formation models have largely been decoupled from discrete choice theory.

In employing discrete choice analysis, we focus on the conditional multinomial logit model, commonly called the conditional logit model for short, which is  a foundational workhorse of discrete choice modeling.
The model belongs to the family of random utility models,
  where choices are interpretable as those of a rational actor selecting
  the alternative with the largest ``utility'' sampled from random variables that decompose into the inherent
  utility of the alternative and a noise term.
With the conditional logit model, we can use existing optimization routines to estimate model parameters and 
existing statistical methods to asses the uncertainty of the estimates.
Discrete choice models can also easily restrict the set
of available alternatives, where it might not be reasonable to assume that the \emph{entire} set of nodes
is available for friendship.
For example, sometimes only ``friends of friends'' are considered~\cite{holme02,jackson07,leskovec08}.

In this paper, we first show that many popular network formation mechanisms can be rewritten
as conditional logit models, including preferential attachment, uniform attachment,
node fitness, latent space models, and models of homophily.
However, the real power of discrete choice models for social network analysis
is the ability to combine different features (e.g., node degree and node age),
as well as different mechanisms (e.g., triadic closure and preferential attachment) and estimate their relative roles.
Social networks are enormously varied in their structure \cite{ikehara17}, but existing methods often do a poor job at modeling this diversity. 
Thus, beyond unifying the network formation and discrete choice literature, we also
develop several new tools for social network analysis.
For example, we show how to estimate models to distinguish the effects of preferential attachment and triadic closure.
We demonstrate these tools by analyzing the formation of the Flickr social network and the formation of a citation network.
We find on Flickr that accounting for triadic closure greatly reduces the estimated role of degree in choosing who to connect to, and that nodes with degree zero have a remarkably high utility.
Our estimates of preferential attachment in the citation network are similar to those observed in prior studies.
When accounting for the age of a paper, we find evidence for linear preferential attachment. However, for a fixed degree, we find that age is \emph{negatively} correlated with the likelihood of a new citation (i.e., older papers are less likely to be cited).

The key assumption underlying our framework is that the available data actually captures edge formation events (either through edge timestamps or other sequential information).
In contrast, many existing approaches to understanding network formation focus on observing only the structural properties of a network at a single point of observation, e.g., its degree distribution, and initiating a deductive process to try and understand how variations in edge formation would lead to different outcomes  \cite{albert99,bianconi01,liu02,jackson07}.
This approach leads to tidy analyses and easy-to-characterize asymptotic properties, but model selection in this context is strongly dependent on what properties are 
compared. Different underlying formation processes can lead to graphs with indistinguishable properties.
For example, many different formation processes result in the same heavy-tailed degree distributions \cite{mitzenmacher03}.
Thus, when ``fitting'' outcome measurements in this way, one has to know (or posit), e.g., the relative rates of node formation and edge formation.
However, when temporal or sequential data is available \cite{holme12,paranjape17},
  our framework overcomes these limitations by incorporating this structure.

\xhdr{Additional related work}
There is a strong connection between our work and work on link prediction and missing data methods using network features to predict edges~\cite{liben07,clauset08}.
A network formation model implicitly makes claims about what edges are most likely to form next, and thus can be evaluated by the same metrics as link prediction algorithms \cite{lu11}.
We use predictive accuracy as a measure of goodness of fit,
but our primary concern is interpretability of the model and estimates, 
which is one of the advantages of the conditional logit model.

In sociology, stochastic actor-oriented models (SAOMs) employ a similar logit choice~\cite{snijders01,snijders10}; 
however, these models are targeted towards data collected as a few snapshots rather than edge-by-edge
formation. SAOMs also model the rate at which nodes form new relationships, whereas we
condition on the node initiating the new edge, providing better estimates of model parameters.
There are also sociological models such as 
relational event models~\cite{butts08} and dynamic network actor models~\cite{stadtfeld17}
that use fine-grained temporal information, yet these also do not condition on the initiator node as we do. 
While these sociological models can incorporate notions of network formation (e.g., preferential attachment),
our conditional logit framework actually cleanly \emph{subsumes} a wide range of models as special cases.

Finally, estimating the parameters that drive edge formation
is different from identifying the factors that could have lead to 
the observed graph. The latter question is often pursued
with so-called exponential random graph models (ERGMs)
\cite{wasserman96,wiuf06,robins07}. 
However, these models do not consider individual edge events,
are hard to estimate, and have known pathologies~\cite{chatterjee13,shalizi13}.

\section{Discrete choice and edge formation}  
\label{sec:model}

We now develop network formation through the lens of discrete choice.
Throughout this paper, we assume that the networks are directed.
Again, while undirected graphs are common in social network analysis,
the actual edge formation process often has directed initiation.
In the common setting of ``growing graphs,'' nodes arrive one at a time
and form edges when arriving in a network. In these cases, the newly
arriving node is considered to be the node initiating the connection;
such analysis is standard with, e.g., classical preferential attachment models~\cite{albert99}.

When modeling the directed formation of an edge $(i,j)$,
  two processes need to be distinguished,
  roughly corresponding to the questions
  ``who is $i$?'' (the chooser) and ``who is $j$?'' (the chosen).
In this paper, we focus on understanding the latter, i.e., the
formation of $(i,j)$ as the selection of $j$ conditional on knowing that $i$
is ready to form an edge.
Thus, our discrete choice models of edge formation can be
readily estimated from data that implicitly or explicitly contains a record of initiating $i$ nodes
and used for subsequent analysis, as we show in Sections~\ref{sec:estimation} and \ref{sec:applications}.
Beyond the scope of this work, our model of ``$j$ conditional on $i$'' can be paired with a model of ``initiations by $i$'' for a full generative model of network formation.

\subsection{Background on discrete choice models}
We now review discrete choice models generically,
which we will then translate to the context of edge formation in Section~\ref{sec:model-logit}.
Consider a universe of alternatives $\mathcal X$ and a dataset 
consisting of $n$ different choices, indexed by $k=1,\ldots,n$.
Each choice $(j, C)$ consists of a choice set $C \subseteq \mathcal X$ and 
a chosen item $j \in C$. The elements in $C$ are mutually exclusive choice alternatives,
and exactly one element from $C$ is chosen.
%For our purposes,
We consider each element $j \in C$ to be represented by a vector of features $x_j$ (for example, in our analysis
in Section~\ref{sec:model-logit}, $x_j$ will be the feature vector of a node in the graph,
and $C$ will represent a set of nodes).
We let $\mathcal D = \{(j_k, C_k)\}_{k=1}^n$ denote the choice data.

A broad family of discrete choice models is the family of random utility 
models (RUMs), of which the conditional logit is an important special case. 
Each alternative $j$ has some inherent utility to the agent $i$ making the choice;
with the conditional logit, we model this utility as a linear function of $j$'s features $x_j$:
\begin{eqnarray*}
u_{i,j} = \theta^T x_j,
\end{eqnarray*}
for some (latent) parameter vector $\theta$ that is fixed across individuals $i$. Random utility models assume that agents make ``rational'' choices by maximizing random utilities
centered on these inherent utilities. 
More formally, the utility $U_{i,j}$ observed by the actor is given by
$U_{i,j} = u_{i,j}  + \varepsilon_{i,j}$,
where $\varepsilon_{i,j}$ is a noise term.
The probability $P_i(j,C)$ of $i$ choosing option $j$ from the choice set $C$ is
\[
P_i(j,C) = \Pr( j = \arg \max_{\ell \in C}\ U_{i,\ell} ).
\]
 
When the $\varepsilon_{i,j}$ are i.i.d. standard Gumbel,~\footnote{Formally, a standard Gumbel distribution is a generalized extreme value distribution with a cumulative density function of the form $F(x) = e^{e^{-x}}$.}
the probability of choosing each alternative 
is proportional to the exponentiated inherent utility \cite{agresti03}:
\begin{eqnarray}
P_i(j,C) = \frac{\exp{\theta^T x_j}}{\sum_{\ell \in C} \exp{\theta^T x_\ell}}.
\label{eq:logit}
\end{eqnarray}

The above model is the conditional multinomial logit model of discrete
choice, though it is often referred to simply as the conditional logit or
multinomial logit (MNL) model.
If the noise terms $\varepsilon_{i,j}$ are distributed i.i.d.~Normal,
then the model is the independent multinomial probit. In probit models,
the choice probabilities are no longer proportional to utilities as in Equation \eqref{eq:logit}.
The Gumbel assumption is common in discrete choice theory
and will facilitate our connections to  a variety
of network formation mechanisms.
There are more complex random utility models that can impose dependence 
between noise terms \cite{train09} and there also is
a growing literature of flexible choice models~\cite{tversky72,benson16,ragain16}
designed to model context effects and other varied 
violations of the {\it independence of irrelevant alternatives} \cite{luce59} (a storied axiom satisfied by the conditional multinomial logit model). We leave the relationships of
these models to network formation as avenues for future research.

\subsection{Edge formation as discrete choice}
\label{sec:model-logit}

With the above formalisms in place, we now develop
network formation from a discrete choice perspective.
We begin by showing how several well-known models
can be conveniently expressed as conditional logit models, with a summary given in
Table \ref{tab:tab1}.
All models are designed to grow simple graphs (i.e., without multi-edges),
and the choice set $C$ excludes any nodes to which the chooser $i$ is already connected.
Every item is represented by its features that, importantly, can evolve over time.
The features $x_{j,t}$ of node $j$ at time $t$ are thus always time-indexed, but we often suppress the $t$ to reduce notational clutter.

\xhdr{Preferential attachment} 
We start with the generalized Barab{\'a}si-Albert model \cite{albert99,krapivsky00,bollobas03},
 also known as the generalized Price model~\cite{price76}, 
one of the most studied models in the network formation literature.
It is typically stated as a growth model of 
a time-evolving graph $G_t = (V_t,E_t)$, $t=1,2,3,\ldots$, and
when a new node arrives it connects to $m$ distinct existing nodes $j$ with a probability
proportional to a power of their degree $d_{j,t}$ at time $t$,
\begin{equation}\label{eq:PA_parametric}
P(j,V_t) = \frac{d_{j,t}^{\alpha}}{\sum_{\ell \in V_t}  d_{\ell,t}^{\alpha}}.
\end{equation}
The exponent parameter $\alpha$ controls the relative
  importance of degree~\cite{krapivsky00}.
The case where $\alpha=1$ is called
  linear preferential attachment,
  and produces networks that can mimic a range of structural properties observed in empirical networks.

If we represent each potential neighbor $j$ with
the time-indexed one-dimensional ``feature vector'' $x_{j,t} = \log d_{j,t}$ and
employ a conditional logit model as in Equation~\eqref{eq:logit},
we obtain a utility
of $j$ for $i$ at time $t$ of $u_{i,j,t} = \theta \log d_{j,t}$. Here
the choice model parameter $\theta$ plays the exact role of $\alpha$,
since $e^{\theta \log d_{j,t}} = d_{j,t}^{\theta}$.

Given a growing network $G_t$, we can construct a choice dataset 
$\mathcal D$ from this network by 
extracting the node $j_t$, node sets $V_t$, and degree sequence
$(d_{1,t}, \ldots, d_{|V_t|,t})$ at each time-step. 
The preferential attachment model has only one parameter, $\theta=\alpha$.
The log-likelihood for that parameter given a dataset is then:
\begin{equation*}
  \begin{split}
    l(\alpha; \mathcal D) 
             &= \sum_{(j,C) \in \mathcal D} \log \frac{\exp(\alpha \log{d_j})}{ \sum_{\ell \in C} \exp(\alpha \log d_\ell)} \\
              &= \sum_{(j,C) \in \mathcal D} \Bigg( \alpha \log{d_j} - \log \sum_{\ell \in C} \exp(\alpha \log{d_\ell}) \Bigg).
  \end{split}
\end{equation*}
We've suppressed the time-index $t$ from the features $\log{d_\ell}$ to reduce clutter, 
but emphasize that $d_\ell$ is the degree at the time of the choice. 

\xhdr{Non-parametric preferential attachment} 
The above model assumes an attachment kernel of a particular parametric form.
From a discrete choice perspective,
one can also estimate the role of degree in edge formation
non-parametrically by estimating a coefficient $\theta_k$
for each degree $k=0,\ldots,n-1$ individually. This approach
has the added benefit of being able to assign 
positive probability to choosing nodes with degree zero.
Under this model, the log-likelihood of the parameters $\theta = (\theta_0,...,\theta_{n-1})$ given the dataset is:
\begin{equation*}
  \begin{split}
    l(\theta; \mathcal D)
              &= \sum_{(j,C) \in \mathcal D} \log \frac{\exp{\theta_{d_j}}}{ \sum_{\ell \in C} \exp{\theta_{d_\ell}}}  \\
              &= \sum_{(j,C) \in \mathcal D} \Bigg( \theta_{d_j} - \log \sum_{\ell \in C} \exp{\theta_{d_\ell}} \Bigg ).
  \end{split}
\end{equation*}
Again we've suppressed time-indexing to simplify the presentation. Pham et al.~\cite{pham15} previously described a version of the above likelihood as a means of measuring the attachment kernel using maximum likelihood, albeit without making the connection to discrete choice.

\begin{table}[tbp]
\caption{Network formation models framed as utility functions for a conditional logit. Where appropriate, we use the traditional notation for the parameters of each process.}
\label{tab:tab1}
\begin{tabular}{l c c}
\toprule
  Process  & $u_{i,j}$ & $C$   \\
  \midrule
  Uniform attachment \cite{callaway01}  & $1$ & $V$   \\
  Preferential attachment \cite{albert99,krapivsky00} & $\alpha \log d_j$ & $V$  \\
  Non-parametric PA \cite{newman01,redner05,pham15} & $\theta_{d_j}$ & $V$  \\
  Triadic closure \cite{rapoport53} & 1 & $\{j: FoF_{i,j} \}$  \\
  FoF attachment  \cite{jin01,vazquez03,saramaki04} & $\alpha\log \eta_{i,j}$ & $V$   \\  
  PA, FoFs only & $\alpha \log d_j $ & $\{ j:FoF_{i,j} \}$   \\
  Individual node fitness \cite{caldarelli02} & $\theta_j$  & $V$  \\ 
  PA with fitness \cite{bianconi01b,motwani06} & $\alpha \log d_j + \theta_j$ & $V$ \\
  Latent space \cite{hoff02,leskovec07,papadopoulos12} & $\beta \cdot d(i,j)$ & $V$ \\
  Stochastic block model \cite{karrer11} & $\omega_{g_i,g_j}$ & $V$ \\
  Homophily  \cite{mcpherson01} & $h\cdot\mathds{1}\{g_i = g_j\}$ & $V$  \\
  \bottomrule
\end{tabular}
\end{table}

\xhdr{Uniform attachment}
A simple edge formation model is to
sample a new neighbor uniformly at random from all nodes~\cite{callaway01}.
There are no parameters in this model, but we can still
write down the likelihood of the model given a dataset, which will be 
useful when we later combine this model with others within
a mixture model:
\begin{equation*}
l(\mathcal D) 
= \sum_{(j,C) \in \mathcal D} \log \frac{\exp{(1)}}{ \sum_{\ell \in C} \exp{(1)}} 
= \sum_{(j,C) \in \mathcal D} - \log{|C|}.
\end{equation*}

\xhdr{Triadic closure}
A variant of uniform attachment is for $i$ to attach to new neighbors uniformly at random
  from the set of their friends-of-friends, as opposed to the 
  set of all nodes. This process effectively models triadic closure~\cite{rapoport53}.
It has the same simple functional form of the uniform model, but now the \textit{choice set}
  $C$ varies with each choice, namely, the choice set is restricted to be only the
  friends of friends of node $i$ (the chooser) to which $i$ is not already connected.
This change in choice set can also be achieved by
assuming the utility of $j$ to $i$ at time $t$ is $u_{i,j,t} = \log ( \mathds{1}\{FoF_{i,j,t}\})$,
where $\mathds{1}\{FoF_{i,j,t}\}$ is a boolean indicating whether $i$ and $j$ are friends of friends at time $t$,
and then letting the choice set revert to the full node set.

An additional model that naturally combines
  the ideas of preferential attachment and
  befriending friends-of-friends
  takes the \textit{number} of friends
  in common between $i$ and $j$ as a feature.
We could define this feature as $\eta_{i,j,t} = |\{ k : e_{i,k,t} \wedge e_{k,j,t}\}|$, where $e_{i,k,t}$ indicates whether there is an edge between $i$ and $k$ at time $t$.
The corresponding utility would be $u_{i,j,t} = \alpha\log \eta_{i,j,t}$.
This model is similar (but not equivalent) to random walk-based formation models \cite{jin01,vazquez03,saramaki04},
  which emphasize formation within a local neighborhood.
 
\xhdr{Node fitness}
Another line of formation models that is subsumed
  by the discrete choice framework are those involving fitness.
In this work, nodes choose to connect to others based on some
  intrinsic latent fitness score.
Certain distributions of fitness values lead to a scale-free degree distribution~\cite{caldarelli02}, 
providing an alternative explanation to preferential attachment for modeling such degree distributions.
We can express the node fitness model by a conditional
logit model with separate fixed effect $\theta_j$ for each node $j$ (so
the feature of a node is an indicator vector of its identity).
The likelihood of the fitness parameters $\theta$ given the data is then:
\begin{equation*}
  \begin{split}
    l(\theta; \mathcal D) 
              &= \sum_{(j,C) \in \mathcal D} \log \frac{\exp{\theta_j}}{ \sum_{\ell \in C} \exp{\theta_{\ell}}}  \\
              &= \sum_{(j,C) \in \mathcal D} \Bigg( \theta_j - \log \sum_{\ell \in C} \exp{\theta_{\ell}} \Bigg ).
  \end{split}
\end{equation*}

This formation model is equivalent to the classic Bradley-Terry-Luce model of discrete choice for estimating the quality of alternatives~\cite{luce59}.
Alternatively, one could replace the individual fixed effects with surrogate features
of node fitness such as an auxiliary measure of
gregariousness (in the case of social networks),
or the impact factor of a paper's journal (in the case of citations networks).
As a simple example from the existing literature, degree and fitness can be easily united in a single ``preferential attachment with fitness'' model \cite{bianconi01}, $u_{i,j} = log d_j + \theta_j$, and remain within the conditional logit framework.

\xhdr{Latent space models}
Another class of network formation models postulates the existence of a latent space
that drives connections between nodes.
Examples of latent spaces include Euclidean space~\cite{hoff02},
hyperbolic space \cite{krioukov2010hyperbolic}, a tree~\cite{leskovec07}, a circle~\cite{papadopoulos12}, or a set of discrete classes~\cite{holland83}.
While the conditional logit model in the form that we describe it does not facilitate finding the best-fitting latent space assignment to explain the data,
  it can be used to estimate the relative importance of a known latent space given
a distance function $d(i,j)$.
As one example from the family of latent space models, in the community-guided attachment (CGA) model~\cite{leskovec07}
  all nodes have a distance derived from the height $h(i,j)$ of common parents in a latent
  tree structure situating all nodes $i$ and $j$.
Given this tree as known, a node connects to another proportionally to $c^{-h(i,j)}$ for some scalar $c>0$.
As a conditional logit model, the corresponding utility function is $u_{i,j} = -h(i,j) \cdot \log(c)$. 
The parameter vector $\theta = \log c$ can be retrieved by fitting a conditional logit with a known
  $h(i,j)$ as the only variable and transforming the estimated
  parameter with $c = \exp(\theta)$. 
Assuming that the latent space representation is given is a strong assumption, and fitting such a model while estimating the latent space representation (e.g.~as done by Hoff et al.~\cite{hoff02} in Euclidean space) is much more difficult.

\xhdr{Additional models}
Conditional logit models are very flexible
  and can deal with multiple features and
  interactions between them.
Any number of features can be added,
  including node covariates and
  structural features like
  a node's clustering coefficient~\cite{bagrow13}
  or age~\cite{callaway01,leskovec08}.
Conditional logit models can also be used to investigate the role
  of homophily \cite{mcpherson01} in edge formation,
  by adding a binary feature indicating whether nodes $i$ and $j$
  are part of the same class.

Table \ref{tab:tab1} summarizes how several network formation models 
  fit within the discrete choice framework via their corresponding utility functions and choice sets.
A major advantage of this framework is that different features
  can easily be combined into a single model and jointly estimated. Or, when suitable, one can employ a mixture of conditional logit models, as we show in the next section.

\subsection{Combining modes using Mixed Logit}
\label{sec:model-mixed}

So far we have written a range of existing and new edge
  formation models as conditional logit models, a specific type of discrete choice model.
But several existing edge formation models that do not fit neatly into the conditional logit framework, meanwhile, align exactly with the use of mixture models in discrete choice modeling. 
Following our success formulating edge formation models as conditional logit models, in this subsection we develop mixed conditional logit formulations of several additional models.

A common proposal to make network formation models more flexible is to augment an existing model by allowing nodes to pick neighbors uniformly at random with some probability $1-p$, while running the ordinary model with probability $p$~\cite{kleinberg99,kumar99,kumar00,krapivsky01,liu02,cooper03}.
This augmentation increases flexibility because it enables the model to explain edge events that may otherwise have probability zero.
Within discrete choice, 
this approach is 
precisely a mixed logit model where one of the mixture modes is uniform attachment. 

While the conditional logit estimates a single parameter vector representing average preferences as shared by all agents, the \emph{mixed logit} model is often used to account for differences in preferences across various types of agents.
In its most general form, the mixed logit is expressed using
  a probability distribution $f$ over
  different instantiations of the parameter vector $\theta$:
$$
P_i(j,C) = \int \frac{\exp{\theta^T x_j}}{\sum_{l \in C} \exp{\theta^T x_l}}  f(\theta) \ d\theta .
$$
In this work, we will only consider discrete mixtures of $M$ logits, also called a latent class model \cite{kamakura89}:
$$
P_i(j,C) = \sum_{m=1}^M  \pi_m \frac{\exp{\theta^T_m x_j}}{\sum_{l \in C} \exp{\theta^T_m x_l}},
$$
where $\sum_{m=1}^M \pi_m = 1$ and the weights $\pi_1,\ldots,\pi_M$ model the relative prevalence of each mode.

\xhdr{Copy model} The copy model~\cite{kleinberg99,kumar99,kumar00} is a classic formation process that can be written as a mixed logit with two modes.
In the first mode,
  new edges connect proportional to degree with probability $p$, 
  while in the second mode they connect uniformly at random with probability 
  $1-p$~\cite{liu02,cooper03}. 
As a conditional logit model, the utilities of the two modes 
are $u^{(1)}_x = \log d_x$ and $u^{(2)}_x = 1$, respectively,
and the class probabilities are $(\pi_1, \pi_2) = (p, 1-p)$.

This model is also called the ``re-direction model'' \cite{krapivsky01}, which is a special case of the copy model when $d$ edges are copied from a sampled vertex; the model here is when $d = 1$, which is often used for analysis~\cite{cooper03,easley2010}. It has also appeared independently elsewhere, constructed as an explicit mixture~\cite{liu02}. The connection between relaxations of preferential attachment and mixture models was also recently
observed by Medina et al.~\cite{medina18}.

\xhdr{Local search model}
Another example of a model with multiple modes is the Jackson-Rogers model of edge formation as a mixture of uniform attachment and triadic closure~\cite{holme02,jackson07}.
The original model is based on a  relative rate $r^{*}$ between edges forming at random and edges formed locally. It also has
  edges form based on respective acceptance probabilities.
We describe a simplified version of this model, which we'll call the local search model, where edges connect to nodes selected uniformly at random from the full node set with probability $r$ and uniformly at random from the set of friends-of-friends with probability $1-r$.\footnote{
Since the $r^{*}$ parameter in the original presentation is actually the \textit{rate} of uniform attachment, we can relate it to our $r$ through $r = \frac{r^{*}}{1+r^{*}}$. For example, if the rate between random and friend-of-friend edges is one to one ($r^{*}=1$), then $r=0.5$.}
We can represent this simplified process with a two-mode mixed logit model.
In this case the mixture parameters are $(\pi_1, \pi_2) = (r, 1-r)$ and
   both modes have the same utility function $u_x = 1$
   but their \textit{choice sets} differ so that the second mode 
   only considers friends-of-friends.\footnote{A model with a restricted choice set, for example to only friends-of-friends, gives a likelihood of zero to choices outside the choice set.}

Table \ref{tab:tab2} overviews the mixture model formulations described above,
as well as a new model---the $(r,p)$-model---that we use in Section~\ref{sec:app2}
to analyze preferential attachment effects.

\begin{table}[tbp]
\caption{Network formation models framed as mixed logits.
Each mixture component is a mode.
FoF refers to friend-of-friending, also called local search or triadic closure.
We define a new $(r,p)$-model as a natural generalization of prior ideas once
we put network formation models in the language of discrete choice.
}
\label{tab:tab2}
\begin{tabular}{l c}
\toprule
  Process  & Modes \\
\midrule
  Copy model \cite{kleinberg99,krapivsky01} & Uniform, PA\\
  Node types \cite{kumar10} & New node, PA, none \\
  Local search \cite{holme02,jackson07} & Uniform, Uniform FoF \\
  ($r,p$)-model  & Uniform, PA, Uniform FoF, PA FoF \\
  \bottomrule
\end{tabular}
\end{table}

\section{Estimation and Inference}
\label{sec:estimation}

To learn a discrete choice model of network formation from data, we assume that we have access to a sequence of directed edges,
  in chronological order. This sequence of edges needs to be recast as choice data in order to fit a choice model.
For every formed edge $(i, j)$, we create a data point consisting of the choice $j$,
the choice set of candidates nodes at the time, and
the features of each candidate node at the time.

Given a data set and a conditional logit model, one can write out the log-likelihood, as shown in Section~\ref{sec:model-logit}.
For any conditional logit model with a linear utility $u_{i,j} = \theta^T x_j$,
the likelihood function is convex with respect to the variables $\theta$
and can be efficiently maximized using standard
  gradient-based optimization (e.g., BFGS).
The functional form of the logit leads to
 straightforward gradients.
For example, for preferential attachment, the gradient is 
\begin{equation*}
  \begin{split}
  \frac{\partial}{\partial \alpha} l(\alpha; \mathcal D)
       &= \sum_{(x,C) \in \mathcal D} \Bigg( \log{d_x} - \frac{\sum_{y \in C} \log{d_y} \cdot \exp(\alpha \log{d_y})}{\sum_{y \in C} \exp(\alpha \log{d_y})}  \Bigg),  \\
  \end{split}
\end{equation*}
where the time-dependence of the features (degrees) have been suppressed to reduce clutter. Gradients for the other choice models in Section~\ref{sec:model-logit} are omitted but straightforward.

One advantage of likelihood-based model fitting is that we can compute standard errors and confidence intervals of the parameters. In particular, the standard errors can be computed with 
$\sqrt{\textbf{H}^{-1}}$~\cite{train09}, where $\textbf{H}$ is the Hessian matrix of second derivatives of the log-likelihood at the parameters.

\xhdr{Mixture models and expectation-maximization}
For mixed conditional logit models, the log-likelihood is no longer
convex in general, making optimization more difficult. 
To maximize the likelihood of mixed models we turn to expectation maximization (EM) techniques \cite{dempster77,train08}.
We briefly summarize the procedure described in Train's book~\cite[Chapter 14.3.2]{train09}.
Assume that we have a model with $M$ modes (i.e., mixture components), 
where every mode starts with initial parameter values $\vec{\theta}^{\,m}$ (usually initiated at 1).
Choices $(x_k,C_k) \in \mathcal D$ are again indexed with $k$, so that $k \in \{1,\ldots,n\}$ and $n = |\mathcal D|$.
The EM algorithm runs through the following steps:

\begin{enumerate}
\item Initiate class probabilities uniformly with $\pi_m = 1/M$ and initial class responsibilities $\gamma^m_k = 1/M$ for each data point.
\item For every data point $k$ and every mode $m$, compute the class responsibility given by the relative individual likelihood: \[\gamma^m_k = \frac{\pi_m \cdot \mathcal{L}^m(\theta^m ; (x_k,C_k))}{\sum_{\ell=1}^M \pi_\ell \cdot \mathcal{L}^\ell(\theta^\ell ; (x_k,C_k))}.\]
\item For every mode $m$, update the total class probability with $\pi_m = \frac{1}{N} \sum_{k = 1}^N \gamma^m_k$.
\item For every mode $m$, update the parameters $\vec{\theta}^{\,m}$ using standard optimization for fitting a single model, weighing each choice set with its class responsibility $\gamma^m_k$.
\item Repeat steps 2--4 until some convergence or stopping criteria.
\end{enumerate}

The total likelihood of the parameters and class probabilities is:
\[l(\theta; \mathcal D) =  \sum_{k = 1}^N \log  \sum_{m=1}^M   \pi_m \cdot \mathcal{L}^m (\theta^m ; (x_k,C_k)) \]
We monitor the convergence of the iterative procedure using the change in this total likelihood between iterations.

Even though EM is theoretically an efficient estimator \cite{xu96},
there are cases when alternatives are appropriate.
For example, if one has reasonable
bounds or priors on the parameter values,
then direct likelihood maximization could be used,
and if the search space is low-dimensional, a grid search might be appropriate. Recent theoretical work has also developed algorithms for learning mixtures of two multinomial logit modes with theoretical guarantees assuming a separation between the modes \cite{chierichetti18}.

\xhdr{Negative sampling}
Every time an edge is formed by some node $i$, 
each node not yet connected to $i$ is a candidate choice.
For large sparse graphs,
  the full choice set of all nodes can become large and the gradients of the log-likelihood expensive to compute.
To speed up this computation, $s$ negative/non-chosen examples can be
  \textit{sampled} uniformly at random to create a (random) reduced
  dataset with smaller choice sets.
For each choice $(j,C)$,
  one forms a smaller random choice set out of the positive choice and the negative samples, 
  $\tilde C \subset C$ with $\lvert \tilde C \rvert =s+1$, and
  replaces the original choice data with $(j,\tilde C)$.
As long as the negative examples are sampled uniformly at random,
  parameter estimates on a dataset with negatively sampled choice sets 
  are unbiased and consistent for the estimates on the 
  on the full set \cite{mcfadden78,train09,jarvis18}.
Practically, there is a trade-off between feature computation and storage on the one hand,
  and the ability to estimate coefficients for rare features on the other.

\begin{figure}
  \centering
  \includegraphics[width=\linewidth]{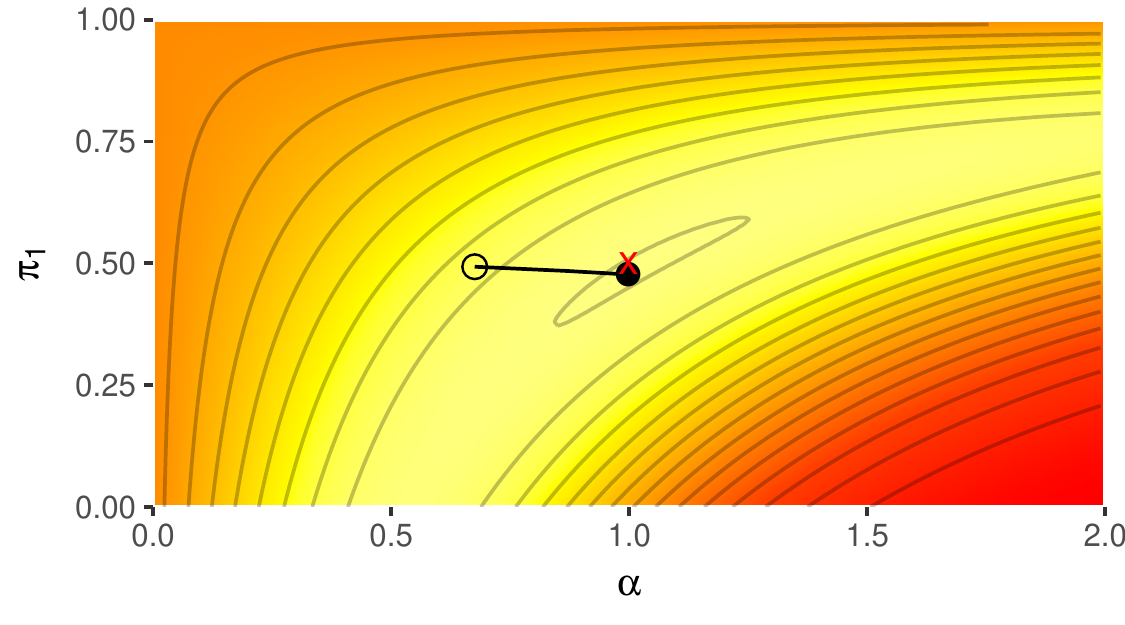}
  \vspace*{-8mm}
  \caption{The log-likelihood surface of the copy model for a graph of $n=10,000$ nodes generated with $\pi_1=0.5$ (marked as x). The line tracks the iterations of the EM algorithm, from open to closed dot.
There is a trade-off between the degree coefficient $\alpha$ and the mixture probability $\pi_1$, but there are large regions with similar likelihoods.}
  \label{fig:fig_ll_surface}
\end{figure}

\xhdr{Typical likelihood surface}
In Figure \ref{fig:fig_ll_surface} we show the representative likelihood surface of a copy model to illustrate its properties.
We generated a synthetic graph on $n=10,000$ nodes according to the copy model with $m=4$ edges per node and degree-attachment probability $\pi_1=0.5$.
We fit a two-mode mixed logit model to this data with $u^{(1)}_j=\alpha \log d_{j,t}$ and $u^{(2)}_j = 1$. We use $s=10$ negative samples. 
There are two free parameters in this model: the degree exponent $\alpha$ and the mixture probability $\pi_1$. We plot the log-likelihood across a reasonable range of values to show that surface is generally well behaved. We see that it is hard to distinguish between
  data generated under a copy model ($\alpha=1$) with probability $\pi_1=0.5$ 
  from data generated from no-mixture ($\pi_1=0$) preferential attachment with $\alpha=0.5$, and there is a general trade-off between the exponent $\alpha$ and the mixture probability $\pi_1$.

\xhdr{Model comparison and the likelihood-ratio test}
Another advantage of our discrete choice framework is that we can employ
standard statistical methods for model selection.
Specifically, when one model is a special case of another, their relative quality can be compared using the likelihood ratio test. 
In the case of the conditional logit,
  a model with additional features can be compared to one without them
  because the latter is a special case of the former
  with the coefficients of the additional features being set to 0.
Or, in the case of the mixed logit,
  one can define a model with multiple modes and manually set some of their class probabilities to zero.

As a concrete example, suppose we wanted to know whether including the age of a node in a preferential attachment model results in a statistically significantly better model.
To do so, we would first estimate the parameters $\theta_1$ of the more complex model, $u^{(1)}_j = \theta_{1,1} \log(d_j) + \theta_{1,2} \log(\text{age})$.
We would then estimate the parameters $\theta_0$ of the simpler model
$u^{(0)}_j = \theta_{0,1} \log(d_j)$.
 Let $\mathcal{L}_1$ and $\mathcal{L}_0$ be the likelihoods of the two models with parameters $\hat \theta_1$ and $\hat \theta_0$.
We can compute the likelihood ratio 
$\lambda = \mathcal{L}_0 /\mathcal{L}_1$.
Under the null hypothesis of the simpler model, with some regularity conditions,
$-2\log\lambda$ is asymptotically distributed $\chi^2_1$ ($\chi^2_k$
where $k$ is the number of additional degrees of freedom in the more complex model)~\cite{wilks38}, a standard test
in the finite regime~\cite[Chapter 3.8.2]{train09}.

\section{Applications}
\label{sec:applications}

We now demonstrate how to use our conditional logit framework to analyze
network formation processes.
We first consider synthetic data and show how our tools
can be used to better analyze preferential attachment mechanisms.
We then analyze two empirical datasets that demonstrate how to
integrate different structural features of the network or integrate node covariates.
In both cases, our framework provides novel insights into the network formation processes.
We provide code for processing data (converting edge lists to choice data) and for model fitting (with negative sampling), available here: \url{https://github.com/janovergoor/choose2grow/}.

\subsection{Measuring preferential attachment}
\label{sec:app1}

\begin{figure}
  \centering
  \includegraphics[width=\linewidth]{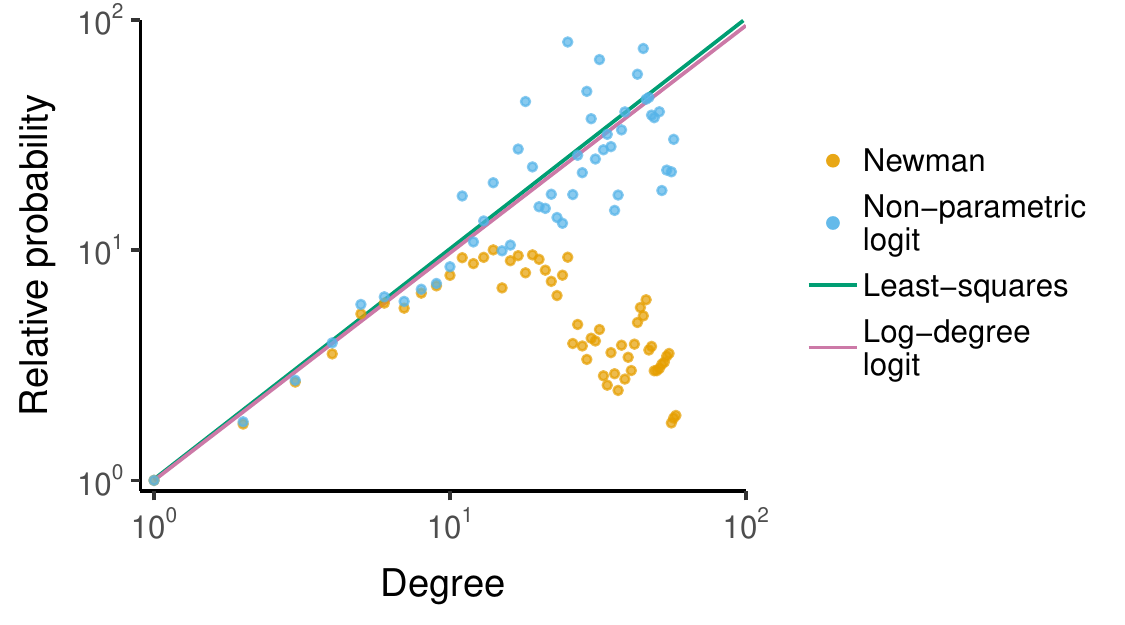}
  \vspace*{-7mm}
  \caption{Attachment kernel fits for a synthetic preferential attachment graph. The Newman measure computes the relative likelihood of selecting a node of that degree, as compared to the likelihood of selecting the lowest degree, but it is biased for higher degrees. The non-parametric logit is consistent but noisy for higher degrees.}
  \label{fig:fig_pham}
\end{figure}

The question of whether and when preferential attachment is an important driver of network formation is widely debated \cite{albert99,callaway01,newman01,caldarelli02,holme02,newman01,vazquez03,saramaki04,jackson07,bagrow13,broido18}.
Most prior research focuses on estimating the shape of the attachment kernel by comparing the degree of chosen nodes to the distribution of available degrees~\cite{newman01,jeong01,redner05}.
However, recent work by Pham et al.\ shows that previous measures are biased~\cite{pham15}.
In particular, the bias comes from the assumption that the
    distribution of available nodes of varying degrees is constant throughout
    the formation process, but this distribution clearly changes
    as the network grows.

To estimate the exponent $\alpha$ of an attachment kernel, Pham et al.~propose fitting something akin to a conditional logit with a separate coefficient for each degree, and then estimating $\alpha$ via a weighted least squares fit over the degree coefficients \cite{pham15}.
Compared to this method, fitting a log-degree logit directly is much simpler.
In fact, it is the maximum likelihood estimator for $\alpha$, and thus consistent and efficient.

To illustrate, we generate a graph with pure preferential attachment ($n=2,000$, $m=1$ edges per node, 
$\alpha=1$) and estimate the attachment kernel by the methods of Newman~\cite{newman01} and Pham et al.~\cite{pham15}.
The maximum degree of this graph was 102, and the results of the different estimation procedures are shown in Figure \ref{fig:fig_pham}.
The non-parametric estimates are similar for lower degrees, but for higher degrees the Newman measure incorrectly drops, illustrating the bias that Pham et al.~have previously documented.
Fitting $\alpha$ directly using a log-degree conditional logit gives an estimate of $\hat \alpha = 0.987$.
The Pham et al.~least squares fit, $\hat \alpha_{LS} = 1.012$, is close to the MLE 
but may deviate considerably in more difficult instances.

\subsection{Disentangling preferential attachment from triadic closure}
\label{sec:app2}

Many models exhibit similar outcomes to preferential attachment \cite{krapivsky00,caldarelli02,holme02,vazquez03,mitzenmacher03,jackson07},
  but there are few principled ways to rigorously test the relative validity of these models.
In this section, we show how to use the discrete choice framework to
  estimate the relative importance of
  preferential attachment while accounting for other dynamics.
To this end, we generate data according to a known generative process
  and fit various (possibly mis-specified) formation models.
Our generative process is a hybrid between
  the copy model of preferential attachment (i.e., choose nodes proportional to degree) and
  the Jackson-Rogers local search model (i.e., connecting to friends-of-friends).
The process, which we call the $(r,p)$-model, is parametrized by
  $r \in (0,1]$ and $p \in (0,1]$.
When a new edge is formed,
  with probability $p$ it is formed uniformly at random
  and with probability $1-p$ it is formed with linear preferential attachment ($\alpha=1$).
Meanwhile, the \textit{choice set} is determined by the second parameter $r$: 
with probability $r$, the choice set is all nodes not yet connected to $i$,
while with probability $1-r$,
the choice set is limited to available friends-of-friends of $i$.
With $r = 1$ this model reduces to the copy-model and with $p=1$ it reduces to the simplified local search model; 
the $(r,p)$-model thus subsumes two popular models in a single, simple discrete choice framework. For a growth process on directed graphs, it is necessary that $p>0$ and $r>0$, otherwise new nodes will never be selected.

\begin{figure}
  \centering
  \includegraphics[width=\linewidth]{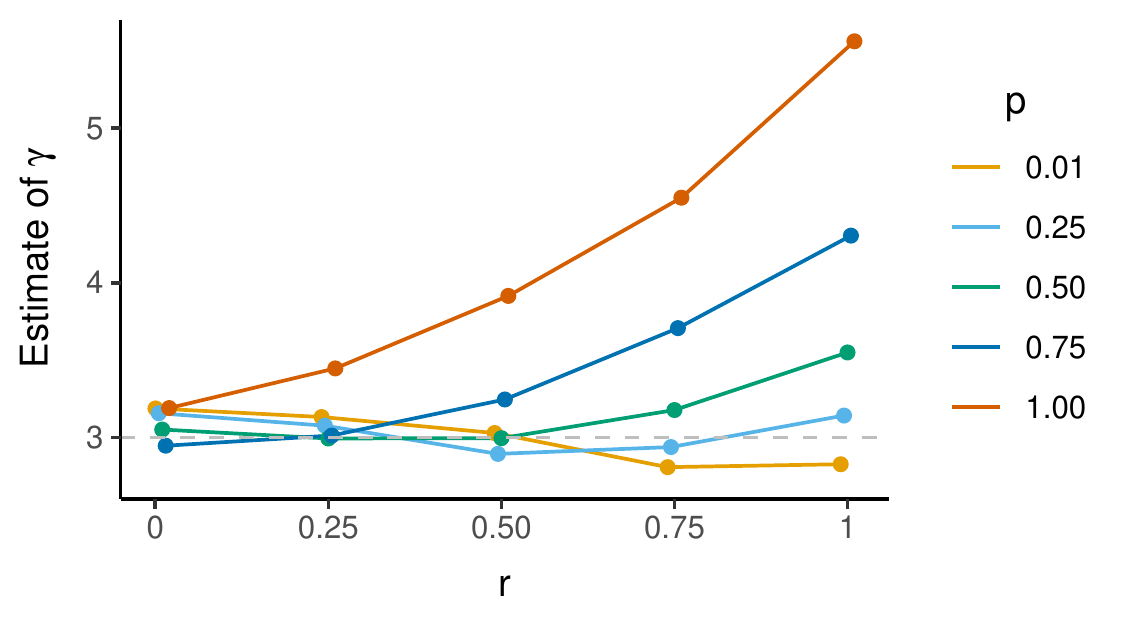}
  \vspace*{-8mm}
  \caption{Estimating the power-law exponent $\gamma$ from the degree distributions of graphs formed under the $(r,p)$-model with $n=20,000$ nodes. Under the local search model with significant triadic closure ($p=1$, small $r$), the exponent looks like it would under the copy model ($p \rightarrow 0$, $r=1$).
 }
  \label{fig:fig_gamma}
\end{figure}

With this general model, we investigate how estimating parameters
of one of the more specific models goes awry when
the true data generating process in fact comes from an instance of the more general model. For a range of values of $p$ and $r$,
  we generated graphs using the following growth process.
New nodes arrive, each creating $m=4$ edges.
For every edge, we sample the mode of the model (according to $r$ and $p$) independently.
If an edge is supposed to be a friend-of-friend edge,
  but no friends-of-friends are available (for example, $i$'s first edge),
  then the process reverts to uniformly random formation across the full node set.\footnote{This creates a slight bias towards uniform at random modes. This reversion to uniform attachment happens for every first edge with probability $1-r$.}
Sweeping through combination of $p$ and $r$ parameter values,
 for each set of parameters 
 we generated 10 \textit{undirected} graphs with $n=20,000$ nodes each.

\xhdr{Degree distributions}
The local search and copy models both produce graphs with power-law degree distributions.
Therefore, fitting a mis-specified model on a degree distribution can lead to misleading results.
To illustrate, we fit a power-law distribution $p(x) \propto x^{-\gamma}$ to the degree distribution of graphs generated from $(r,p)$-models using maximum likelihood estimation~\cite{clauset09},
with estimates for $\gamma$ in Figure~\ref{fig:fig_gamma}.
In theory, an undirected graph formed with the copy model process with probability parameter $p$ leads to a degree distribution with power law exponent $\gamma=(3-p)/(1-p)$ \cite{mitzenmacher03,bollobas03} 
(for directed graphs, $\gamma=(2-p)/(1-p)$).
As $p$ increases, the degree distribution looks more like a random graph without preferential attachment. However, as $r$ goes down (increasing the relative role of friend-of-friends), the parameter estimate looks like the estimates for the copy model, even when $p=1$.

To summarize, it is not recommended to estimate a formation model from an observed degree distribution.
The parameter estimates are sensitive to small deviations in the generative process.

\begin{figure}
  \centering
  \includegraphics[width=\linewidth]{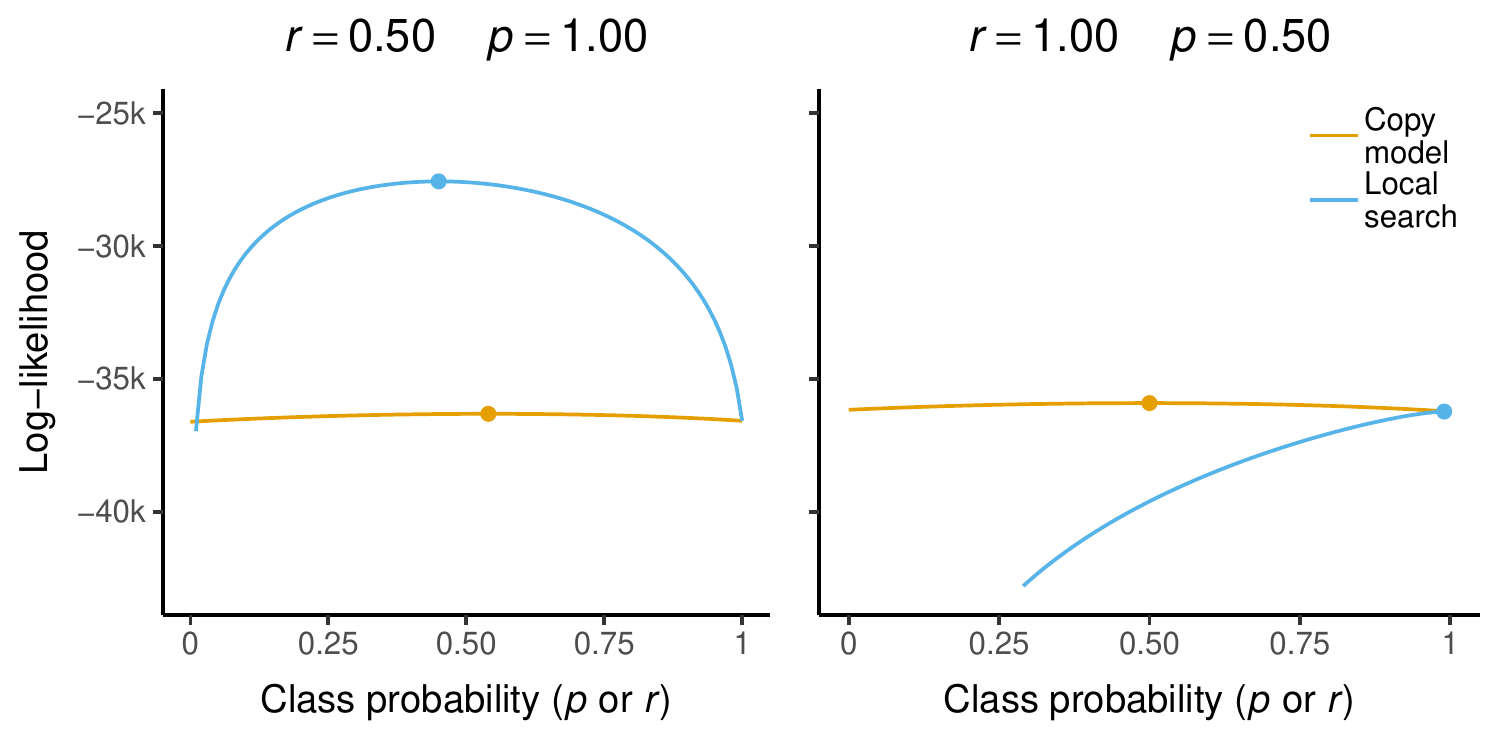}
  \vspace*{-7mm}
  \caption{The log-likelihood of varying the class probabilities of the copy model ($r=1$, $p$ free) or the local search model ($r$ free, $p=1$) for two different synthetic graphs. In both cases the true model is the most likely. On the left we see a large difference in the log-likelihood between optima, while on the right we see a smaller difference. In both cases a likelihood ratio test is highly significant ($P$-values $< 10^{-16}$).
  }
  \label{fig:fig_ll_compare}	
\end{figure}

\xhdr{Discrete choice modeling}
Beyond degree distributions, in Figure~\ref{fig:fig_ll_compare} we look at how the two subsumed models (the copy model and the Jackson-Rogers local search model) fare when estimated from formation data generated by the $(r,p)$-model. We look at two cases. As a first case, we generate graphs with $r=0.5$ and $p=1$, so half the edges are formed to friends-of-friends with no utility from degree. The likelihood under a local search model ($r$ free, $p=1$) as a mixed logit is maximized at $r=0.45$, while for the copy model ($r=1$, $p$ free) it is maximized at $p=0.54$. The former is a much better fit than the latter ($P$-value < $10^{-16}$), and the copy model erroneously thinks that preferential attachment is driving 45\% of the edges.
As a second case, we look at a graph generated with $r=1$ and $p=0.5$, so half the edges are due to preferential attachment, and friend-of-friending plays no role. In this case, both models are correctly maximized at their relative values. Again, the correct model has a higher likelihood ($P$-value < $10^{-16}$).

\subsection{Choosing to follow on Flickr}
\label{sec:app3}

We now apply our framework to examine a real-world network formation dataset capturing the growth of the Flickr social network.
We find that incorporating a Friend-of-Friend feature beyond
preferential attachment and link-reciprocation features 
substantially improves both likelihood and test accuracy and furthermore
that the inclusion of this feature significantly reduces
preference for degree-based attachment.
However, omitting preferential attachment entirely leads to a worse model.
We also find a preference for nodes with zero degree over low degree nodes.
This hints that such nodes play a special role in the network formation process,
even though they would be ignored in preferential attachment models.

\xhdr{Data}
We use a scrape of the Flickr social network collected daily between October 2006 and May 2007~\cite{mislove07,mislove08}.
Users of Flickr can choose to follow other users and the ``following'' (but not the ``followed by'') connections are publicly accessible.
The data was gathered using a breadth-first search crawl, which means that only the connected components reachable from the seed profiles are represented in the data.
Since a full crawl was performed daily,
  the timing of new edges can be identified at the granularity of a day.
The graph contains 3.2 million nodes and 33.1 million edges.

As described in the original papers, this data is consistent with both
  preferential attachment, as inferred from the in-degree distribution, and
  local search, as inferred from the over-representation of edges to nodes that are close to the linking node \cite{mislove08}.
Fitting a power law to the distribution of \textit{in}-degrees gives $\hat \gamma = 1.741$, which would indicate super-linear preferential attachment.
We can test the relative importance of triadic closure by
  fitting a Jackson-Rogers model using the degree distribution matching procedure described in \cite{jackson07}.
This results in $\hat{r}=0.252$, estimating that three out of four edges are formed through triadic closure.

\begin{table}[tbp]
  \caption{Conditional logit model fits for Flickr data.
  Standard errors of the estimates are given in parentheses.
  Evaluation statistics are computed over 2,000 sampled examples excluded from the training data. 
 }
  \label{tab:flickr}
\centering 
\tabcolsep=0.3cm
\begin{tabular}{lcccc} 
\toprule
& \multicolumn{4}{c}{Model} \\
\cmidrule{2-5} 
& \#1 & \#2 & \#3 & \#4 \\ 
\midrule  
log Followers  &  1.149*   &             &  0.715*    &  0.536*    \\  
               &   (0.007) &             &   (0.009)  &  (0.010)   \\  
Has degree     & -0.580*   &             &  -0.631*   & -1.745*    \\
               &   (0.202) &             &   (0.190)  &  (0.234)   \\
Reciprocal     &  8.419*   &   8.347*    &  8.197*    &  7.903*    \\  
               &   (0.220) &    (0.220)  &   (0.240)  &  (0.244)   \\  
Is FoF         &           &   6.12*    &  3.955*    &            \\  
               &           &    (0.045)  &   (0.050)  &            \\  
2 Hops         &           &             &            &  6.290*  \\  
               &           &             &            &  (0.190)   \\  
3 Hops         &           &             &            &  2.851*  \\  
               &           &             &            &  (0.185)   \\  
4 Hops         &           &             &            &  0.583*  \\  
               &           &             &            &  (0.189)   \\  
5 Hops         &           &             &            & -0.585*  \\  
               &           &             &            &  (0.218)   \\  
$\ge$ 6 Hops      &           &             &            & -1.122*  \\  
               &           &             &            &  (0.266)   \\
\midrule 
Observations   &   20,000  &    20,000   &   20,000   &   20,000   \\
Log-likelihood &  -16,448  &   -14,685   &  -10,728   &   -9,789   \\
Test accuracy  &    0.758  &     0.722   &    0.853   &    0.855   \\
\midrule \\[-1.8ex]
\textit{Note:}  & \multicolumn{4}{r}{*p$<$0.01} \\
\bottomrule
\end{tabular} 
\end{table}

\xhdr{Discrete choice analysis}
We fit a series of conditional logit models to further investigate the network formation process.
We isolated a sample of 20,000 edge formation events occurring around the same date,\footnote{We enumerated edges starting November 5, 2006 and included new edges with probability 0.01 until reaching the desired sample size. 
We excluded edge events originating from nodes seen for the first time in a given day (the timing of these edge events are uncertain due to the original data collection process).
The same analysis starting on March 3, 2007 led to virtually identical results.}
  to avoid time heterogeneity affecting the estimates.
We fit several models, displayed in Table \ref{tab:flickr}.
Not-chosen alternatives are negatively sampled with $s=24$.
We log-transform in-degree (representing the number of followers),
  but in order to account for nodes with degree zero, we add a ``has degree'' feature for having a positive degree and use a modified version of $\log$ that returns 0 for input 0.\footnote{This solution is better than using $\log d+\epsilon$, or giving degree-zero nodes the same utility as degree-one nodes. Either of those solutions will give substantially different results, especially when there are many degree-zero nodes.}
In the first column, we fit a model using just these two degree-related features,
  and a reciprocity feature capturing whether the target node is already following the chooser.
Reciprocity is a common phenomenon, with 60\% of edges being followed back \cite{mislove08}.
The estimate $\hat \alpha$ (the coefficient for ``log Followers'') for this model is significantly larger than 1, again consistent with super-linear preferential attachment. 

In the second model, we test the effect of the target node being a friend-of-friend of the choosing node.
In the case of Flickr, this means that the choosing user already follows someone that follows the target user, which evidently is strongly correlated with following that user.
However, combining these two features in a third model (column 3) leads to both estimated parameters dropping substantially.
Most remarkable is the 40\% drop in the estimate of $\alpha$,
  which paints a very different picture about the role of degree.

In the fourth model,
  we measure network proximity as in the original paper,
  by counting the number of ``hops'' (path length) from $i$ to the target before an edge was made.
We integrate the hops as categorical variables to show the relative impact of each additional ``hop''.
Being two hops away is equivalent to being a friend-of-friend, and thus has strongly positive coefficient.
Every additional hop corresponds to a sharp decrease in choosing that node.
Being five hops away is slightly worse than there not being a path at all. 
This could be an artifact of the way the data was gathered,
  so that new regions of the graph only get ``discovered'' when there is at least one link to them,
  or this could be due to path length not being an accurate measure of distance for newer nodes.
Since the number of hops is co-linear with being a friend-of-friend, we can't test them both at the same time.

\begin{figure}
  \centering
  \includegraphics[width=\linewidth]{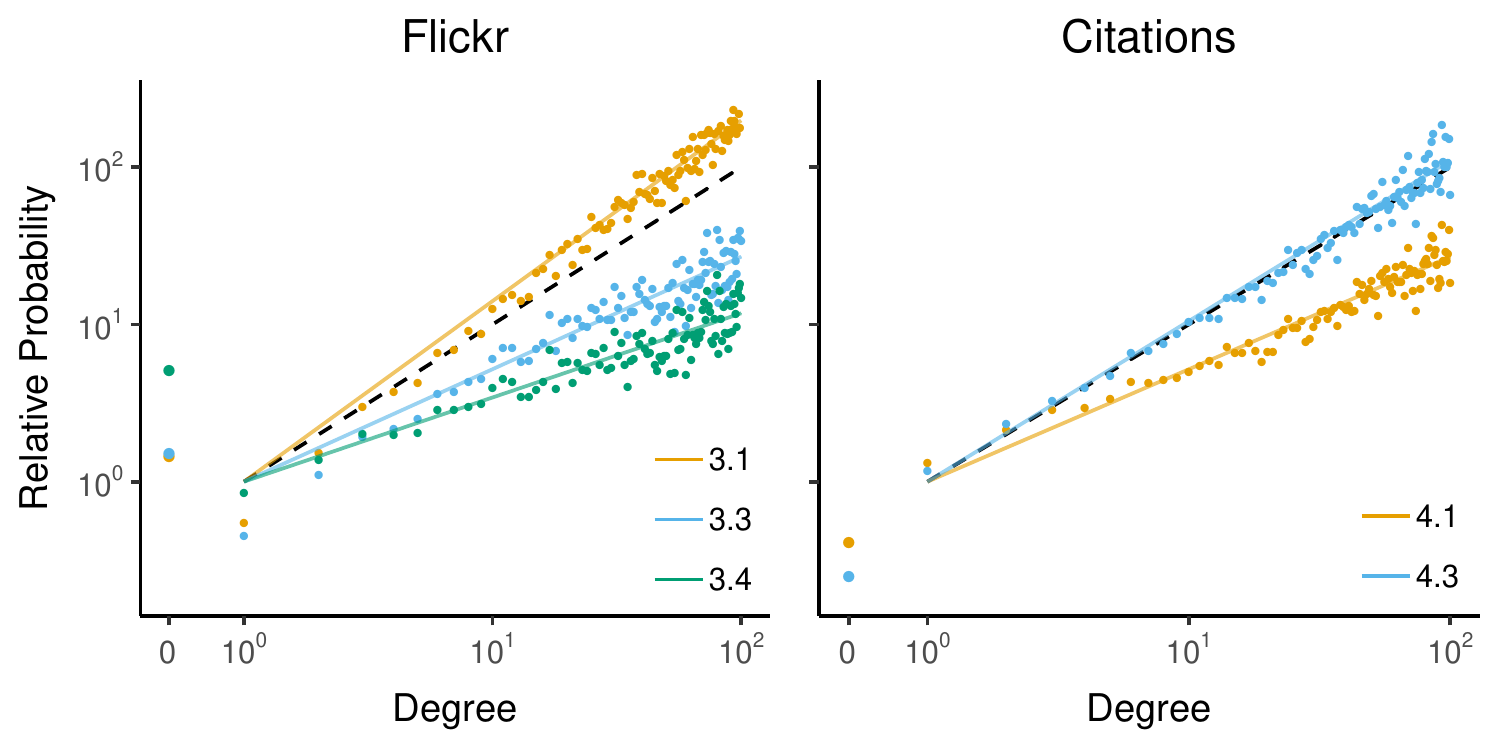}
  \vspace*{-7mm}
  \caption{The probability of being chosen by degree, as compared to a node with degree 1. We show the fits of parametric (lines) and non-parametric (points) conditional logit models of the Flickr and citation networks. The legend references model numbers in Table~\ref{tab:flickr} and Table~\ref{tab:mag}. The estimate for degree 0 is inserted for comparison. Dashed reference lines illustrate what exact linear preferential attachment ($\alpha=1$) would look like.}
  \label{fig:fig_applications}
\end{figure}

In Figure \ref{fig:fig_applications} we visually show the effect of different specifications on the estimate of $\hat{\alpha}$. The first model of the Flickr data looks like super-linear preferential attachment, while the role of degree in the other two is significantly reduced. However, fitting a non-parametric model shows that the estimated coefficients for individual degrees are remarkably linear, suggesting that the functional form of $d_j^{\alpha}$ is a good fit for this network.
One important point is the role of zero-degree nodes.
In most descriptions of preferential attachment, nodes with degree zero are not considered.
However, in the Flickr data set, zero-degree nodes have a higher utility than positive low degree nodes, which could again be an artifact of the data collection process, or point to the special role of new nodes in the network.
Either way, our framework allows one to find these kinds of patterns, and investigate them further.

\subsection{Choosing to cite}
\label{sec:app4}
We now turn to citation network data to show 
  how a discrete choice framework
  facilitates the testing of network formation hypotheses.
Previous analyses of citation networks have observed
 linear preferential attachment with respect to degree \cite{redner05} and
 bias towards citing more recent work \cite{redner05}.
 Here, we find consistent results that older papers are less likely to be cited
 but that accounting for age actually \emph{increases} the importance of
 degree (i.e., after accounting for age, higher degree nodes are more likely
 to be cited).

\xhdr{Data} 
We use the Microsoft Academic Graph\footnote{The Aminer Project \cite{tang08,sinha15}, \url{https://aminer.org/open-academic-graph}}
dataset and focus on a representative subgraph of 459,000 ``Climatology'' papers.
We focus on the subgraph of a single field to simplify the analysis
  since citations are predominantly within the same field of study
  (our analysis was similar on other subgraphs).
We construct a graph out of this data by adding an
  edge each time a paper in our dataset cites another paper
  in our dataset.
For our analysis of Climatology publications, 45\% of edges are within the domain and 
citations to papers that are not labeled are excluded, leaving 3 million edges.
We sample 10,000 citation events uniformly at random from papers published after 2010 and apply negative sampling ($s=24$).
This processing results in 10,000 choices with 25 alternatives in each choice set.
For each possible choice, we compute four features:
  the number of citations at the time of citation,
  whether the paper shares authors with the citing paper,
  the age of the paper in years at the time of citation,
  and the maximum number of publications by any one of the authors at the time of publication. This last feature is a proxy for node fitness \cite{caldarelli02}.

\xhdr{Discrete choice analysis}
We fit conditional logit choice models relating
  these features to the likelihood of citation (Table \ref{tab:mag}).
The first model (first column) is a simple log-degree model.
We find that the estimate $\hat \alpha$ (the coefficient for ``log Citations'') is substantially lower than one,
  consistent with sub-linear preferential attachment.
Apart from the log-likelihood of the models, we also report the predictive accuracy
  (defined as the share of instances predicted correctly)
  on a holdout test set of 2,000 examples.
Just relying on prior degree already gives an accuracy of 36\%,
  which is high for a classification task with 25 classes.
In model two (second column), we add a covariate for whether a paper shares an
  author with the citing paper. As expected, this has a strongly positive coefficient.
 
 \begin{table}[tbp]
  \caption{Learned conditional logits for the ``Climatology'' citation network.
  Standard errors of the estimates are given in parentheses.
  Evaluation statistics are computed over 2,000 sampled examples excluded from the training data. 
 }
  \label{tab:mag}
\centering 
\tabcolsep=0.2cm
\begin{tabular}{lcccc} 
\toprule
& \multicolumn{4}{c}{Model} \\
\cmidrule{2-5} 
& \#1 & \#2 & \#3 & \#4 \\ 
\midrule 
log Citations          & 0.717*   &  0.794*    &  1.052*     & 1.044*     \\
                       &  (0.008) &   (0.010)  &   (0.012)   &  (0.012)     \\
Has degree             & 1.684*   &  1.677*    & 1.862*      & 1.830*     \\
                       &  (0.053) &  (0.062)   &  (0.063)    &  (0.064)     \\
Has same author        &          &  6.523*    &  5.928*     & 5.913*     \\
                       &          &   (0.110)  &   (0.114)   &  (0.114)     \\
log Age                &          &            &  -1.096*    & -1.069*    \\
                       &          &            &   (0.018)   &  (0.021)     \\
Max papers by author   &          &            &             & 0.029*    \\
                       &          &            &             &   (0.011)    \\
\midrule  
Observations           & 10,000   &  10,000    &   10,000    &  10,000  \\
Log-likelihood         &-20,799   & -16,600    & -14,384     & -14,390  \\
Test accuracy          & 0.358    &   0.484    & 0.533       & 0.534    \\
\midrule \\[-1.8ex]
\textit{Note:}  & \multicolumn{4}{r}{*p$<$0.01} \\
\bottomrule
\end{tabular} 
\end{table}

For the third model we add a covariate for the age of the paper in log years (years is always at least one).
Older papers are less likely to get cited (accounting for degree), 
  but accounting for age \textit{increases} the relative importance of degree significantly.
This expanded model also increases the accuracy to 53\%,
  indicating that these feature weights
  do capture substantially more predictive power.
Finally, in model four we add the ``max papers by authors'' feature as a proxy for fitness. The coefficient is small but positive. Accounting for fitness slightly reduces the estimated relative importance of degree, but the $\hat \alpha$ estimate is still close to 1. Adding this feature does not improve the log-likelihood or predictive accuracy; a better proxy for fitness may explain the data better.
Looking back to the visual display of $\alpha$ for the citation models in Figure \ref{fig:fig_applications}, the non-parametric coefficients are highly linear.
In this data, zero-degree nodes are significantly \textit{less} attractive than nodes with degree one. 

As with any regression, the identifying causal effects
  from model fit depends on the design of the study.
The estimates we provide here,
  as is the case with most analyses of observational data,
  are descriptive and not meant to describe causal processes.
The point is that discrete choice models provide a flexible framework
  to easily test and compare different hypotheses around network formation.

\section{Discussion}

When modeling network formation, the majority of the literature analyzes networks that grow ``externally,'' with new nodes arriving and choosing who to connect to, and this setting has also been our main focus here. External growth leads to convenient models that are relatively easy to analyze, with citation networks and patent networks as examples of empirical networks that follow this generative process reasonably closely.
However, in many (especially social) networks, pairs of older nodes often form edges as well, edges that are ``internal'' to the existing set of nodes. 
An extreme example is the social networks of schools or classrooms, which have a fixed node population and ``grow'' purely through an internal growth process.
A major advantage of modeling network formation as discrete choice is that it does not require any model of edge event initiation and simply conditions on the sequence of decisions to initiate, focusing the modeling on the choices made by the initiator. Discrete choice can therefore easily be used to model internal growth as well.

Another major advantage of discrete choice modeling is that it connects the analysis of large-scale network datasets to statistical methods (fitting generalized linear models) that are tremendously scalable. As we show in this work, additional techniques (e.g.,~negative sampling) makes it possible to efficiently scale the estimation process to very large network datasets.

Since the conditional logit model of discrete choice is a random utility model,
  the estimated parameters can be interpreted as the marginal utility of each feature.
This allows one to question the functional form of features.
For example, we show that preferential attachment is equivalent to the logarithmic utility of degree.
Given that degree is commonly heavy-tailed, this is a natural functional form, but we point out that the conditional logit allows one to flexibly compare different specifications.

Our discrete choice perspective has implications for how network data is best collected and analyzed.
It is useful to consider and record notions of directionality, even if edges can otherwise be considered to be undirected.
With information about the choice set associated with each choice, we can see what each node $j$ looked like at the time the choice was made.
Datasets that record the exact time of all edge formation events, as opposed to lumping edge events at the granularity of days or years,  makes it possible to further analyze the formation process in more detail.

There are a couple limitations to our proposed methodology.
First, we cannot model purely undirected edges without some notion of direction.
Second, even though the conditional logit and mixed logit models allow one to model
  similar mechanisms, the interpretations of their estimates are different.
The estimates of a conditional logit are more akin to those of a linear regression model, where one estimates the expected change in an outcome from varying a covariate.
A mixture model is a probabilistic combination of constituent modes,
  so the class probabilities indicate the relative importance to each mode, which makes it harder to compare the roles of individual features within or across modes. However, many traditional models of network formation are equivalent to mixture models, which motivated our consideration of them in this work. 
  
By making foundational connections between network formation and discrete choice, we are hopeful that many further tools from discrete choice theory can be applied to the study of network formation.
For example, there can be bias in network formation, e.g., men are more likely to cite themselves than women \cite{king17}. Our discrete choice framework can help study these cases more rigorously.
For another example, discrete choice models of subset selection \cite{fishburn96,benson18} could be applied to understand possible substitution and complementarity effects in network formation.
And discrete choice interpretations of machine learning embeddings techniques \cite{rudolph16} can likely help unpack the behavior of recent embedding-based network representation methods such as DeepWalk \cite{perozzi14}.
Networks fundamentally represent interactions between discrete entities, and it is therefore natural that methods for modeling and analyzing discrete choice should enable many contributions.

\xhdr{Acknowledgements} We thank Aaron Clauset, Eduardo Laguna-M{\"u}ggenburg, and Daniel Larremore for their helpful comments and feedback.
ARB was supported by NSF Award DMS-1830274
and ARO award W911NF-19-1-0057.
JO and JU were supported in part by an ARO Young Investigator Award.

\bibliographystyle{ACM-Reference-Format}
\balance 
\bibliography{choosing_to_grow}
\end{document}